\documentclass[11pt]{article}
\usepackage{sao1}
\usepackage{psfig}

\advance\textheight by\topskip 

\begin{document}

\title{On the Photometric Redshift Estimates for FR\,II Radio Galaxies
\footnote{
Astronomy Letters, Vol. 31, No. 4, 2005, pp. 219-224.
Original Russian Text Copyright 2005
by {\it Verkhodanov, Kopylov, Pariiskii, Soboleva, Temirova}.
Translated from Pisma v Astronomicheskii Zhurnal,
Vol.{\bf 31}, No.4, 2005, pp.243-249.
}}

\author{
O. V. Verkhodanov\footnote{e-mail: vo@sao.ru} \and
A. I.  Kopylov                                \and
Yu. N. Pariiskii                              \and
N. S.  Soboleva                               \and
A. V.  Temirova
}

\institute{
\saoname
}
\date{September 13, 2004}{}
\maketitle

\begin{abstract}
Using the photometric data on FR II radio galaxies obtained in the Big
Trio Program and
data from other sources, we confirmed the stable correlation between
the spectroscopic and photometric
redshifts up to $z\sim4$ determined from the evolutionary synthetic
spectra of elliptical galaxies. This is a
confirmation for the theoretical predictions of the existence of a stellar
population at high redshifts and its
subsequent evolution corresponding to the population of giant elliptical
galaxies.

.
\end{abstract}

2005 Pleiades Publishing, Inc.

\section{Introduction}

The photometric redshifts that were first used by 
Baum back in 1962 (Baum 1962) still remain an 
important tool for studying the distant Universe (see, 
e.g., Botzler et al. 2004; Rocca-Volmerange et al. 
2004; Vanzella et al. 2004; Budavari et al. 2003). 
Being measured from the spectral energy distribution, 
they are of paramount importance as the next-to-last 
selection step in searching for distant objects (Bunker 
et al. 2003; Franx et al. 2003; Idzi et al. 2004), especially with
the emergence of full sky surveys such as
SDSS (Abazajian et al. 2004) where the photometric 
classification is also used (Padmanabhan et al. 2004). 
The photometric zbecome a more powerful selection 
factor in searching for distant radio galaxies in combination with
the radio-astronomical selection based
on the spectral indices and morphology (Pariiskii et
al. 1996). 

Morphological signatures of radio galaxies, namely, 
type-II (FR II) objects in the classification by Fanaroff and Riley (1974),
help to reveal distant objects
for several reasons: 

(1) these are intense and relatively young (compared to FR I) radio
sources visible at any $z$;

(2) they have a formed radio structure, which is 
circumstantial evidence that the source has a reserve 
of time for the formation of its stellar population; 

(3) the FR II radio galaxies have more intense 
emission lines than the FR I galaxies (Baum et al. 
1995), which facilitates their spectroscopic studies. 

In general, relatively close ($z<2$) FR\,II radio
galaxies are identified with giant elliptical galaxies 
with high radio luminosities and old, homogeneous 
stellar populations. Therefore, this type of objects 
is convenient to use for photometric studies in radio cosmology
(Pariiskii 2001). Note that the age and
composition of the stellar population in radio galaxies can also be
determined from the absorption lines
and the continuum formed by this stellar population. 
However, the sensitivity is insufficient to detect such 
lines in the spectra of many distant radio galaxies.
Besides, the measurement and identification procedure
with the determination of the correct weight for each 
line isfairlycomplex. Using the continuum makes it
possible to apply our knowledge about the composition and age
of a radio galaxy's stellar population
based on evolutionary models. 

Currently available models predict the fast formation (within 1\,Gyr)
of such systems even at
$z\sim4$ (Pipino and Matteucci 2004), which allows
photometric methods to be used to study them. 
The efficiency of selecting such galaxies by radio-
astronomical methods, starting from moderate red 
shifts ($z>0.5$), was confirmed by several teams
(Pedani 2003). The combined Hubble $K-z$ diagram
for radio galaxies and field galaxies (Jarvis et al. 
2001; De Breuck et al. 2002) shows that the radio 
galaxies have the highest luminosities at any redshift, 
0 <z<5.2 (Reuland et al. 2003). In addition, the 
radio galaxies have supermassive black holes whose 
mass is generally proportional to that of the stellar 
bulge ($M_{BH}\sim0.006M_{bulge}$; Magorrian et al. 1998),
which is further evidence for the presence of a formed 
stellar population. The formation of radio galaxies 
at redshifts $z\sim3-5$ gives the already formed stellar
populations at $z\sim2-4$ in the $\Lambda$CDM models. Thus,
when selecting distant radio galaxies, we effectively 

\begin{figure}
\centerline{
\hbox{
\psfig{figure=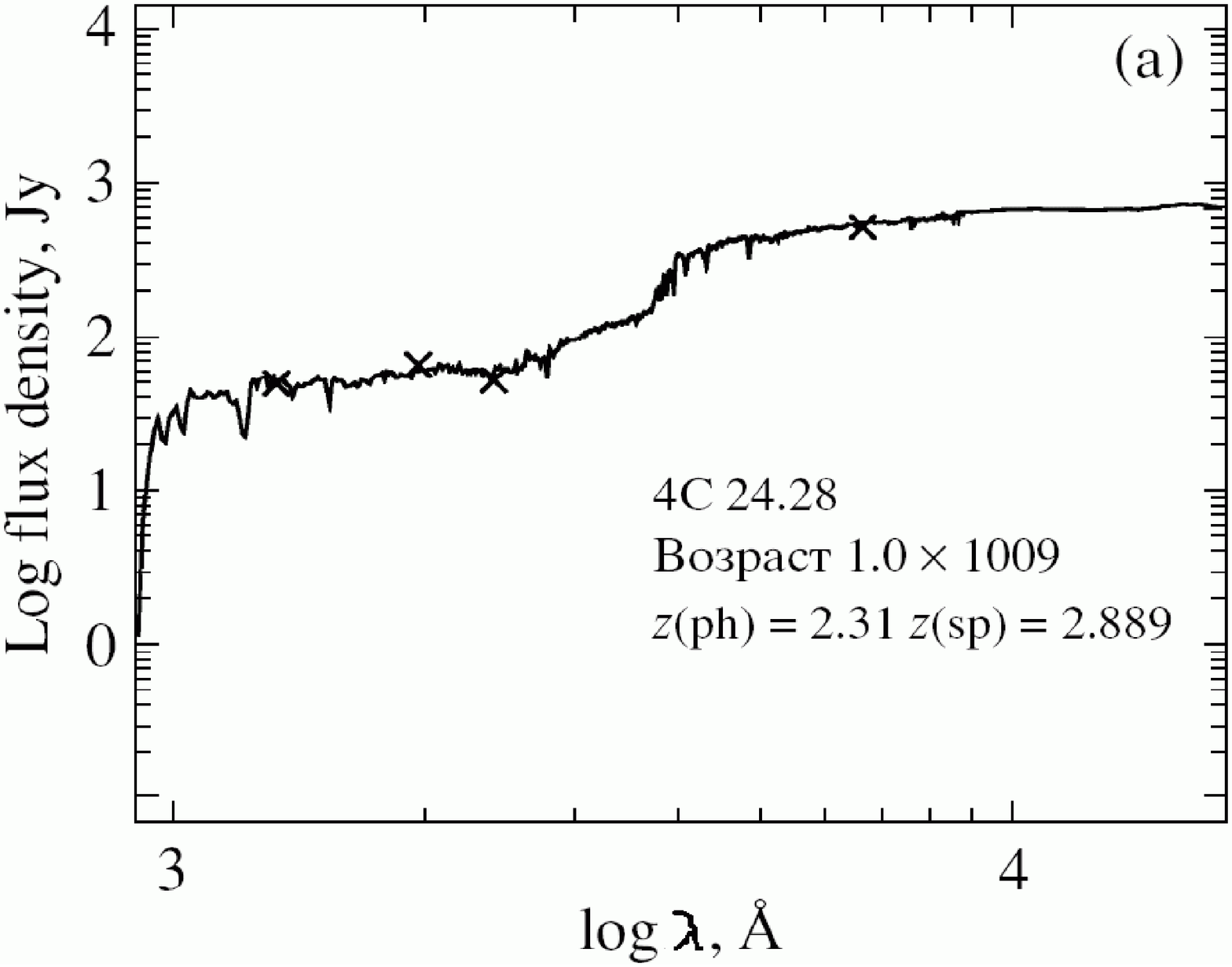,width=8cm,angle=-90}
\psfig{figure=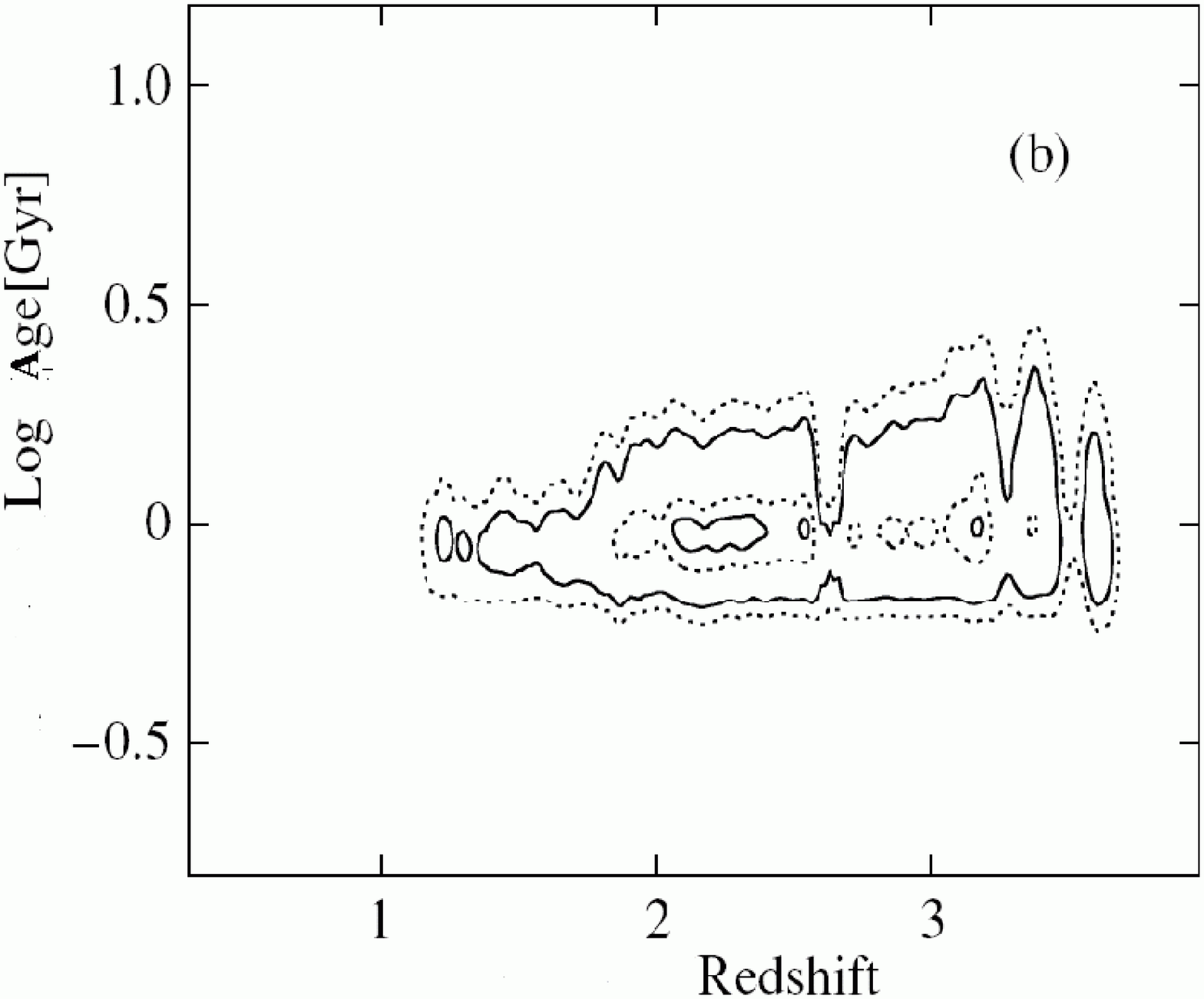,width=7.5cm,angle=-90}
}}
\caption{
Demonstration of determining the photometric redshift for the radio
galaxy 4C\,24.28: (a) the two-dimensional section of
the three-dimensional likelihood function that depends on the redshift,
age, and flux density; (b) the positions of the photometric
points in the SED curve corresponding to the maximum of the likelihood
function.
}
\end{figure}

single out giant elliptical galaxies that can be used 
for photometric studies. The above factors show the 
possibility of using the photometric technique to 
determine the redshifts from the stellar population for 
FR II radio galaxies at $z>2$.

The most recent theoretical calculations (see, e.g., 
Rocca-Volmerange et al. 2004) indicate that evolutionary models of
the spectra for elliptical galaxies can
be applied even to radio galaxies with $z\sim4$, when
gE galaxies with a mass of $10^{12}M_{\sun}$ have already been
formed. 

Nevertheless, the presence of intense emission 
lines (e.g., H$\alpha$, C IV, He II) in the spectra of several
distance radio galaxies that were formed by various 
ionization mechanisms (Maxfield et al. 2002) near 
jets and ``hot spots'' raises the question of whether the
evolutionary spectra of elliptical galaxies correspond 
to those of radio galaxies and, as a result, whether the 
photometric technique can be used to estimate the 
redshifts of radio galaxies. 

In this paper, we consider the $z_{phot}-z_{sp}$ relation
for two samples of FR\,II radio galaxies that we
drew in our previous papers (Pariiskii et al. 1996;
Verkhodanov et al. 1999, 2000a, 2000b) using the 
popular PEGASE (Projet d'Etude des Galaxies par
Synthese Evolutive) evolutionary model of the stellar population
(Fioc and Rocca-Volmerange 1997;
Le Borgne and Rocca-Volmerange 2002). Apart 
from this model, we can use the GISSEL (Galaxy 
Isochrone Synthesis Spectral Evolution Library) 
model (Bruzual and Charlot 1993, 1996; Bolzonella 
et al. 2000), which yields similar results for the photometric redshifts.
The files with the spectral energy
distributions (synthetic spectra) for both evolutionary 
models of the stellar population can be found at 
{\tt http://sed.sao.ru} (Verkhodanov et al. 2000).

\section{The redshift estimation procedure}

We took into account the absorption in our Galaxy 
using maps from Schlegel et al. (1998) written in the 
form of FITS files. 

Before using the model curves, we smoothed 
them with filters by applying the following algorithm 
(Verkhodanov et al. 2002): 
$$
S_{ik} = \frac{\sum\limits_{j=1}^n s_{i-n/2+j}f_{ij}(z)}
	   {\sum\limits_{j=1}^n f_{ij}(z)}
 \eqno(1)
$$
where $s_i$ is the initial model SED curve, $S_{ik}$ is
the model SED curve smoothed with the $k$-th filter,
$f_k(z)$ is the transmission curve of the $k$-th filter
``compressed'' by a factor of $(1 + z)$
when ``moving''
along the wavelength axis of the SED curve, and $j$
is the point number in the filter transmission curve. 
We constructed the two-dimensional ($\lambda$, filter) array
of smoothed synthetic stellar spectra from the $k$
SED curves formed in this way for the subsequent 
calculations. 

We estimated the redshifts and ages of the stellar populations
by choosing the optimal positions
of the photometric magnitudes obtained in various 
bands during the observations of radio galaxies in 
the smoothed SED curves. We used the computed 
and tabulated SED curves for different ages. The 
algorithm of choosing the optimal positions of the 
data points in the curve (Verkhodanov et al. 1996) 
\begin{figure}
\centerline{\psfig{figure=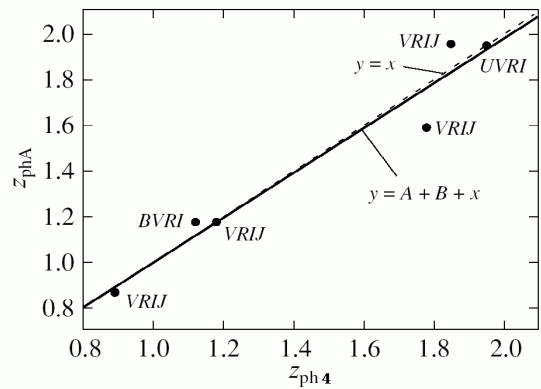,width=7.5cm,angle=-90}}
\caption{
The ``photometric redshift based on all colors
($z_{phA}$) --- photometric redshift based on four colors
($z_{ph4}$)''
relation. The dotted line indicates the ordinate --- abscissa
($x=y$) straight line. The solid line represents the $z_{phA}-z_{ph4}$
regression with the coefficients $A=0.0098$ and $B=0.98531$.
}
\end{figure}
consisted in shifting the observational points along 
the wavelength and intensity axes of the SED curves. 
In this way, we found the position at which the sum 
of the squares of the deviations of the data points 
from the corresponding smoothed curves was at a 
minimum; i.e., we actually calculated the minimum 
of the $\Xi^2$ value:
$$
\Xi^2 = \sum\limits_{k=1}^{Nfilters}\left(\frac{F_{obs,k}-pSED_k(z)}
		  {\sigma_k} \right)\,
 \eqno(2)
$$
where $F_{obs,k}$ is the observed magnitude in the kth
filter, $SED_k(z)$ is the model magnitude for a given
spectral energy distribution in the $k$-th filter at given
$z$, $p$ is a free coefficient, and $\sigma_k$ is the measurement
error. The redshift was determined from the shift in 
the positions of the observed magnitudes at their best 
positions in the SED curves from the ``rest frame'' position.
From the whole set of curves for different ages,
we chose those for which the sum of the squares of the 
residuals for the observational data of radio galaxies 
was at a minimum. Figure 1 shows an example of 
applying the described technique to the radio galaxy 
4C\,24.28 or, more specifically, the two-dimensional
section of the three-dimensional likelihood function
dependent on the following parameters: the redshift,
the age, and the flux density, along with the positions
of the photometric points in the SED curve
corresponding to the maximum of the likelihood function.

We checked the validity of our redshift (and age) 
estimates by two methods.
First, we took the synthetic spectra obtained by smoothing the SED curves
for different ages with filters. This procedure enabled 
us to model the CCD observations for five filters. 
Subsequently, we chose the data points corresponding to the VIJHK filters,
for example, at the redshift

\begin{figure}
\centerline{\psfig{figure=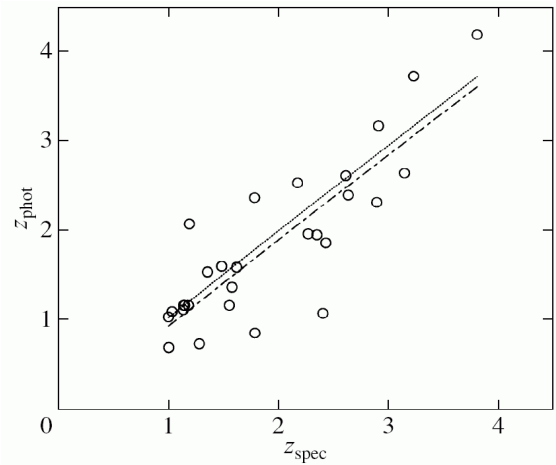,width=7.5cm,angle=-90}}
\caption{
The ``spectroscopic redshift --- photometric redshift''
relation for FR\,II radio galaxies with $z>1$.The
dot---dashed straight line represents the regression based
on all of our data (the correlation coefficient is 0.85). The 
dotted straight line represents the regression with the 
data for the objects 3C\,239, 3C\,266, 4C\,34.34 excluded
(the correlation coefficient is 0.91). 
}
\end{figure}
$z=0.54$, and the evolutionary spectra with ages of 1
and 5\,Gyr.
For each age, we performed two magnitude estimation tests:
with fixed $z=0.54$ and unfixed
redshift. These tests lead us to conclude that both the 
age and the redshift are determined reliably. However, 
there is a possibility of falling on the neighboring 
age curve, which gives an error of 200\,Myr, and the
result for unfixed $z$ is also affected by the quantization
in wavelength $\lambda$ in the SED curves (the error in $z$
reaches 6\%).

In the second case, we explored the possibilities 
of determining the redshifts and ages of the stellar 

\begin{figure}
\centerline{\psfig{figure=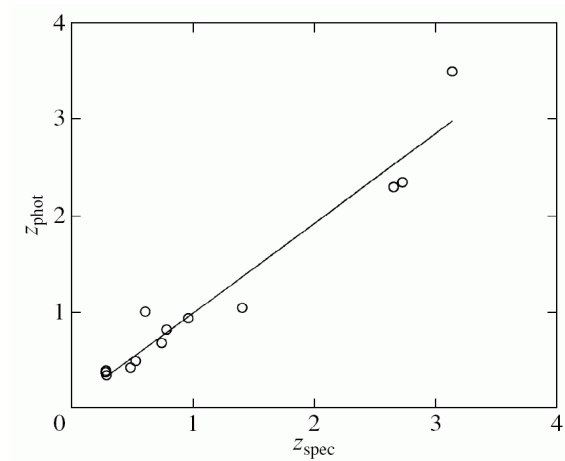,width=7.5cm,angle=-90}}
\caption{
The ``spectroscopic redshift --- photometric redshift''
relation for the radio galaxies of the ``Big Trio''
Program. 
}
\end{figure}

\begin{table}[!t]
\caption{
The photometric redshifts for distant FR II galaxies estimated
from published photometric data (Verkhodanov et al. 1999)
The table gives the radio galaxy's name, its spectroscopically measured
redshift, and its photometric $z$ estimate.
}
\begin{tabular}{l|l|l}
\hline
Name          &$z_{spec}$  &  $z_{phot}$      \\
\hline
3C 13         &  1.35      &  1.54            \\
3C 65         &  1.176     &  1.17            \\
3C 68.2       &  1.575     &  1.37            \\
MRC 0316-257  &  3.142     &  2.64            \\
MRC 0406-244SE&  2.427     &  1.86            \\
4C 41.17      &  3.8       &  4.20            \\
3C 184        &  0.994     &  1.04            \\
3C 194        &  1.185     &  2.07            \\
3C 239        &  1.781     &  0.86            \\
3C 241        &  1.617     &  1.59            \\
B21056+39     &  2.171     &  2.53            \\
4C 34.34      &  2.40      &  1.08            \\
3C 266        &  1.275     &  0.74            \\
3C 267        &  1.142     &  1.17            \\
4C 26.38S     &  2.609     &  2.61            \\
4C 39.37      &  3.225     &  3.73            \\
3C 280        &  0.996     &  0.70            \\
B21256+36     &  1.13      &  1.12            \\
4C 24.28      &  2.889     &  2.31            \\
3C 294        &  1.779     &  2.36            \\
53W091        &  1.55      &  1.17            \\
3C 368        &  1.132     &  1.17            \\
4C 40.36      &  2.267     &  1.96            \\
4C 48.48      &  2.348     &  1.95            \\
MRC 2025-218  &  2.63      &  2.39            \\
3C 437        &  1.48      &  1.60            \\
4C 28.58      &  2.905     &  3.17            \\
PK 2353-018   &  1.028     &  1.10            \\
\hline
\end{tabular}
\end{table}

populations in the parent galaxies from multicolor 
photometry. For this purpose, we selected about 40 
distant radio galaxies with known redshifts and with 
published magnitudes at least in three filters
(Verkhodanov et al. 1998b, 1999). First, we determined only
the ages of the stellar populations of the parent galaxies
at fixed known redshifts from the collected
photometric data. Then, we searched for the optimal model
SED curve with the simultaneous determination of 
the redshift and the age of the stellar population. 
Subsequently, we compared the values obtained.
Using this method, we estimated both the galaxy's age
and its redshift based on the given models (see also 
Verkhodanov et al. 1998a, 1999). It is clear from general considerations
that the reliability of the result at
high redshifts depends significantly on the availability 
of infrared data (up to the Kband), since our fitcovers 
the range of fast change in the spectrum (a jump) 
before the optical SED range; thus, we can determine 
the position of our data reliably, with a pronounced 
maximum in the likelihood curve. Indeed, in testing 
the reliability of our procedure using the available 
measurements when keeping only three points, one 
of which is in the Kband , we obtain the same result 
in the likelihood function as that obtained from four 
or five points. If, however, the infrared range is not 
used, then the result is found to be more uncertain. 
However, as we showed previously (Verkhodanov et 
al. 1999), the case of four closely spaced filters, as in 
our case of BVRIphotometry, yields a good result in 
the check sample of six objects (Fig. 2) that matches 
the result obtained by using all filters, including the 
infrared band. Note that this set of filters works stably at medium
redshifts ($z~1$), which closely corresponds to the sample of radio
galaxies in the ``Big Trio'' Program.

\section{The samples of objects}

\subsection{Radio Galaxies with $z>1$}

As was mentioned above, to test our technique and 
to estimate the redshifts and ages of stellar systems, 
we drew our sample of distant ($z>1$) FRII radio galaxies with
spectral indices $\alpha<-1.0$ and with
redshifts up to $z=3.80$ (Verkhodanov et al. 1998b,
1999) from the data obtained by other authors. 

It should be noted that the published photometric 
data are highly inhomogeneous. They were obtained 
not only by different authors, but also on different 
instruments and with different filters.
The measurements for the same object were not always made
with the same apertures, etc. Therefore, after the final selection,
only 42 of the 300 radio galaxies from
the primary sample remained. Most of the objects 
were not included in the sample, because they have 
the properties of quasars, which severely complicates 
the use of the SED procedure for standard elliptical 
galaxies. From the remaining objects, we chose only 
the radio galaxies with an FR II structure (28 galaxies, Table 1). 

Figure 3 shows the ``spectroscopic redshift ---
photometric redshift'' relation for FR II radio galaxies
with $z>1$. The correlation coefficient calculated from

\begin{table}[!t]
\caption{
The redshifts and their photometric estimates for the objects of
The table gives the object's name, its spectral index, the ratio of
the flux densities at 1.4 and 3.9\,GHz, the R-band magnitude,
the spectroscopic and photometric redshifts, and the radio
source's morphological type.
the RC catalog studied in the ``Big Trio'' Program.
}
\begin{tabular}{cccclll}
\hline
Name    & $\alpha$ & $S_{1400}/S_{3900}$ & $R$ & $z_{sp}$ &$z_{ph}$& Notes \\
\hline
J0105+0501 &  1.05   &     79/25    & 22.8 & 3.14  &  3.5  & FR II ?      \\
J0444+0501 &  1.09   &    214/69    & 22.7 & 2.73 :&  2.35 & FR II        \\
J0209+0501 &  1.16   &     89/33    & 18.5 & 0.285 &  0.38 & Pointlike    \\
J0457+0452 &  1.12   &    201/56    & 19.4 & 0.482 &  0.41 & FR I         \\
J0908+0451 &  0.92   &    301/109   & 19.6 & 0.525 &  0.48 & FR II        \\
J1124+0456 &  0.94   &    935/400   & 17.8 & 0.284 &  0.36 & FR II        \\
J1155+0444 &  1.0    &    141/54    & 18.6 & 0.289 &  0.33 & FR II        \\
J1333+0451 &  1.3    &     42/11    & 18.1 & 1.405 &  1.04 & FR II QSR    \\
J1339+0445 &  1.07   &    119/41    & 22.6 & 0.74  &  0.67 & FR II Triple \\
J1626+0448 &  1.26   &    191/46    & 22.9 & 2.656 &  2.30 & FR II        \\
J1722+0442 &  0.99   &    763/300   & 20.7 & 0.604 &  1.0  & FR II        \\
J2029+0456 &  0.69   &    142/53    & 21.7 & 0.78  &  0.81 & FR II        \\
J2224+0513 &  0.93   &    346/107   & 21.3 & 0.96  &  0.93 & FR II        \\
\hline
\end{tabular}
\end{table}

all of the data in Table 1 is 0.85. If we discard the 
outlying data attributable to random errors, where 
$z_{spec}$ is much higher than $z_{phot}$ (the objects 3C\,239,
3C\,266, 4C\,34.34), then the correlation coefficient
becomes 0.91. 

\subsection{Radio Galaxies from the RC Catalog}

We drew our second sample from FR\,II radio
galaxies with steep spectra discovered in
the RATAN--600 ``Kholod'' survey (Parijskij et al. 1991, 1992)
using multicolor photometry to estimate the color 
redshifts and the ages of the parent galaxies' stellar
systems (Parijskij et al. 1996; Verkhodanov et al.
2002). In the ``Big Trio'' observations
(Parijskij et al. 1996, 1998), the {\it BVRI} magnitudes were measured
for about 60 radio galaxies, and it was found that 
although their color ages had a large dispersion, the 
redshifts could be estimated reliably. 

Subsequently, 20 objects were observed
spectroscopically with the SCORPIO instrument at the
6\,m BTA telescope (Afanasiev et al. 2002, 2003).
Their $z$ measurements (Kopylov et al. 1995a, 1995b;
Parijskij et al. 1996, 1998) and photometric estimates
based on BTA observations (Parijskij et al. 1996; 
Verkhodanov et al. 2002a, 2002b) were obtained 
before spectroscopy by Afanasiev et al. (2002, 2003) 
and are collected in Table\,2.

As we see from Table\,2, the $z$ measurements confirm our photometric
estimates (Fig.\,4) even for some
of the quasars. The correlation coefficient estimated 
from the observations of Kholod radio galaxies is 0.92. 

\section{Conclusions}

Using the photometric data for FR\,II radio galaxies with steep
spectra obtained as part of
the ``Big Trio'' Program and data from other sources in the
astronomical literature, we have confirmed the stable 
correlation between the spectroscopic and photometric redshifts up to
$z\sim4$ determined from the evolutionary synthetic spectra of
elliptical galaxies. The
technique that we have used since 1995 (Parijskij et
al. 1996) confirms its efficiency in selecting distant 
radio galaxies and investigating their evolutionary 
properties. 

The fact that the photometric estimates for FR\,II
radio galaxies are close to direct measurements suggests
that the continuum optical spectra of radio
galaxies can be described by stellar models, while 
the evolutionary models of elliptical galaxies used 
correspond to the stellar population of distant radio 
galaxies and are generally evidence of its existence at 
high redshifts. This conclusion allows us to further 
use the photometric methods based on evolutionary 
models with a high degree of confidence to study FR II 
radio galaxies and to estimate cosmological parameters
(Verkhodanov and Parijskij 2003; Starobinsky et al. 2004).

\section{Acknowledgements}

O.V. Verkhodanov and A.I. Kopylov are grateful
to the Russian Foundation for Basic Research
(project no. 02-07-90038) for partial support of this 
work. Yu.N. Pariiskii was supported by grants from
the ``Integration'' and ``Astronomy''
Programs and the
Russian Foundation for Basic Research. We wish to 
thank the referee for remarks that improved the paper.

Translated by N. Samus'

\end{document}